 \documentclass[5p,times,twocolumn]{elsarticle}

\usepackage{amsmath,amsfonts,amsthm,amssymb}
\usepackage{lipsum}
\usepackage{physics}
\usepackage{nicefrac}

\usepackage{yfonts}

\biboptions{comma,sort&compress}
\usepackage[utf8]{inputenc}

\usepackage{bm}
\usepackage{hyperref} 
\usepackage[dvipsnames]{xcolor}
\usepackage{nicefrac}
\newcommand{\pvec}{{\boldsymbol p}}

\hypersetup{
colorlinks = true,
linkcolor = blue,
anchorcolor = blue,
citecolor = blue,
filecolor = blue,
urlcolor = blue
}
\usepackage{ulem}

\journal{Physics Letters B}

\begin{document}

\begin{frontmatter}



\title{Longitudinal spin polarization in a thermal model with dissipative corrections}


\author[a]{Soham Banerjee}
\ead{soham.banerjee@niser.ac.in}
\affiliation[a]{organization={School of Physical Sciences, National Institute of Science Education and Research, An OCC of Homi Bhabha National Institute},
            city={Jatni},
            postcode={752050}, 
            country={India}}
            
\author[b]{Samapan Bhadury}
\ead{samapan.bhadury@uj.edu.pl}
\affiliation[b]{organization={Institute of Theoretical Physics, Jagiellonian University},
            city={Kraków},
            postcode={30-348}, 
            country={Poland}}   
            
\author[b]{Wojciech Florkowski}
\ead{wojciech.florkowski@uj.edu.pl}

\author[a,b]{Amaresh Jaiswal}
\ead{a.jaiswal@niser.ac.in}

\author[c]{Radoslaw Ryblewski}
\ead{radoslaw.ryblewski@ifj.edu.pl}
\affiliation[c]{organization={Institute of Nuclear Physics, Polish Academy of Sciences},
            city={Kraków},
            postcode={31-342}, 
            country={Poland}}              
\begin{abstract}
In this work, we address the problem of longitudinal spin polarization of the $\Lambda$ hyperons produced in relativistic heavy-ion collisions. We combine a relativistic kinetic-theory framework that includes spin degrees of freedom treated in a classical way with the freeze-out parametrization used in previous investigations. The use of the kinetic theory allows us to incorporate dissipative corrections (due to the thermal shear and gradients of thermal vorticity) into the Pauli-Lubanski vector that determines spin polarization and can be directly compared with the experimental data. As in earlier similar studies, it turns out that a successful description of data can only be achieved with additional assumptions --  in our case, they involve the use of projected thermal vorticity and a suitably adjusted time for spin relaxation ($\tau_s$). From our analysis, we find that $\tau_s \sim 5$ fm/$c$, which is comparable with other estimates.
\end{abstract}


\end{frontmatter}

\section{Introduction}
\label{introduction}

One of the recent motivations for studying non-central heavy-ion collisions is to study the spin polarization effects of various produced particles, for example, $\Lambda$ hyperons~\cite{STAR:2017ckg, STAR:2018gyt, STAR:2019erd} and vector mesons~\cite{ALICE:2019aid}, for a recent review see Ref.~\cite{Niida:2024ntm} and for various related theoretical developments see Refs.~\cite{Liang:2004ph, Liang:2004xn, Becattini:2009wh, Montenegro:2017rbu, Montenegro:2017lvf, Florkowski:2017ruc, Florkowski:2017dyn, Florkowski:2018ahw, Sun:2018bjl, Hattori:2019lfp, Xie:2019jun, Florkowski:2019qdp, Weickgenannt:2019dks, Weickgenannt:2020aaf, Bhadury:2020cop, Shi:2020htn, Singh:2020rht, Yi:2021ryh, Wang:2021ngp, Hongo:2021ona, Florkowski:2021wvk, Wang:2021wqq, Hongo:2022izs, Weickgenannt:2022zxs, Sarwar:2022yzs, Wagner:2022amr, Kumar:2022ylt, Biswas:2023qsw, Weickgenannt:2023btk, Kiamari:2023fbe, Kumar:2023ojl}. In particular, there is a lot of interest in the measurement of the longitudinal spin polarization of $\Lambda$'s produced in Au-Au collisions at the beam energy of $\sqrt{s_{\rm NN}} = 200$~GeV \cite{STAR:2019erd}. In this case, the data indicates a quadrupole structure of the longitudinal polarization in the transverse momentum plane, which disagrees in sign with most theoretical predictions (the so-called sign problem)~\cite{Becattini:2017gcx}.

In this paper, we continue some of the earlier studies of the longitudinal polarization~\cite{Florkowski:2019voj, Florkowski:2021xvy} that use thermal model to parametrize freeze-out conditions in Au-Au collisions. A novel feature of the current analysis is that we incorporate dissipative corrections to the Pauli-Lubanski vector. They are determined within the framework of kinetic theory, which treats spin degrees of freedom classically~\cite{Florkowski:2018fap}. This approach turns out to be consistent with the formalism that uses Wigner functions in the semiclassical expansion and considers GLW versions of the energy-momentum and spin tensors.

The most common approach in the polarization studies is to assume that the spin polarization tensor $\omega_{\mu\nu}$ is given by the thermal vorticity  
\begin{equation}
\varpi_{\mu\nu}=-\frac{1}{2}(\partial_\mu \beta_\nu - \partial_\nu \beta_\mu).
\end{equation}
Here $\beta^\mu$ is the ratio of the hydrodynamic flow and temperature, $\beta^\mu = u^\mu/T$. This leads to the expression for the Pauli-Lubanski vector that includes the spin equilibrium distribution function depending on $\varpi_{\mu\nu}$. The kinetic-theory approach allows for including corrections to the equilibrium distributions. In our case, they are obtained from the kinetic equation treated in the relaxation time approximation (RTA)~\cite{Bhadury:2020puc, Bhadury:2020cop, Bhadury:2022ulr}. In this way, we naturally obtain contributions from the thermal shear
\begin{equation}
    \xi_{\mu\nu}=\frac{1}{2}(\partial_\mu \beta_\nu + \partial_\nu \beta_\mu)
\end{equation}
and gradients of thermal vorticity. We note that the effects due to the thermal shear were first considered in Refs.~\cite{Fu:2021pok, Becattini:2021suc, Becattini:2021iol}. Thus, our approach offers yet another method to incorporate such corrections.

We use the freeze-out parametrization introduced in Ref.~\cite{Broniowski:2002wp}. It turned out to be very successful in describing many observables like the transverse-momentum spectra or elliptic flow coefficients of various hadronic species. However, as the model is boost-invariant, it cannot reproduce the global spin polarization that points in the out-of-plane direction. On the other hand, as it correctly reproduces the elliptic flow, it is natural to use it in the studies of longitudinal polarization as these two phenomena are considered to be interconnected~\cite{Niida:2024ntm}.

It has been demonstrated before that inclusion of the thermal shear helps to overcome the sign problem~\cite{Fu:2021pok, Becattini:2021suc, Becattini:2021iol}. However, further conjectures were usually made to reproduce the data (for example, neglecting temperature gradients at freeze-out~\cite{Becattini:2021iol} or exchanging the $\Lambda$ mass by the strange quark mass~\cite{Fu:2021pok}). In our case, we also introduce an additional assumption -- it involves the use of projected thermal vorticity. By projected vorticity we mean the thermal vorticity tensor with all ``electric'' like components neglected, i.e., with $\varpi_{0i} = 0$. As found in our previous work~\cite{Florkowski:2019voj}, this procedure is consistent with a non-relativistic treatment where polarization is determined solely by the spatial components of the rotation $\partial_i v_j - \partial_j v_i$ \cite{Voloshin_2018}.

The use of the projected thermal vorticity leads to a correct sign of the quadrupole structure of the longitudinal spin polarization, however, there remains a quantitative difference between theoretical predictions and the data~\cite{Florkowski:2019qdp}. In this work, we show that this disagreement can be removed by the inclusion of dissipative corrections described above with a suitably fitted spin relaxation time. Our numerical calculations show that a very good description of the effect can be obtained with the spin relaxation time $\tau_s \sim 5$ fm/$c$, which is in agreement with earlier estimates of the spin relaxation time~\cite{Kapusta:2019sad, Ayala:2020ndx, Hidaka:2023oze, Wagner:2024fhf}.

\smallskip
{\it Notation and conventions}: For the Levi-Civita tensor
$\epsilon^{\mu\nu\alpha\beta}$ we follow the convention $\epsilon^{0123} =-\epsilon_{0123} = +1$. The metric tensor is of the form $g_{\mu\nu} = \textrm{diag}(+1,-1,-1,-1)$. Throughout the text we make use of natural units, $\hbar = c = k_B = 1$.


\section{Spin observables at freeze-out}
The mean spin polarization of $\Lambda$ hyperons can be found from the Pauli-Lubanski (PL) vector~\cite{Florkowski:2017dyn, Florkowski:2018fap}
\begin{equation}
E_p \frac{d \Delta \Pi_\alpha (x,p)}{d^3p} =-\frac{1}{2} \epsilon_{\alpha \mu \nu \beta } \, \Delta \Sigma_\lambda E_p\frac{dS^{\lambda,\mu \nu}}{d^3p}\frac{p^\beta}{m},
\label{eq:PL1}
\end{equation}
where $p^\mu = (p^0, \pvec)$ is the $\Lambda$ four-momentum with the on-mass-shell energy $p^0 = E_p = \sqrt{m^2 + \pvec^2}$, $m$ is the $\Lambda$ mass ($m = 1.116$~GeV), $\Delta \Sigma$ is an infinitesimal element of the freeze-out hypersurface, and $S^{\lambda,\mu \nu}$ is the spin tensor discussed in more detail in \ref{appA}. Integration of Eq.~(\ref{eq:PL1}) over the freeze-out hypersurface gives the momentum density of the PL vector
\begin{align}
    E_p \frac{d \Pi_\tau (p)}{d^3p} &= - \frac{1}{2} \frac{\cosh \xi}{(2\pi)^3 m}  \int \Delta \Sigma \cdot p\, e^{-\beta \cdot p}\, \epsilon_{\tau \mu \nu \beta } p^\beta\nonumber\\
    &\quad \times \left[\left(1 + \chi \right) \omega^{\mu \nu }  - \frac{\tau_s}{u \cdot p}\;  p \cdot  \partial\, \omega^{\mu \nu } \right] . \label{PL-KT}
\end{align}
Here $\xi$ is defined as the ratio of the baryon chemical potential and temperature, $\omega^{\mu\nu}$ is the spin polarization tensor, $\tau_s$ is the spin relaxation time, and $\chi= [\tau_s/(u \cdot p)] \, p^\rho p^\sigma \xi_{\sigma \rho}$.

As we are interested in the spin polarization per particle, we introduce the particle number density
\begin{equation}
    E_{p} \frac{d N(p)}{d^3p} = \frac{4 \cosh\xi}{(2\pi)^3} \int  \Delta \Sigma \cdot p\, e^{-\beta \cdot p}.
\end{equation}
A quantity which can be directly compared with the experimental results is the ratio of the integrated densities
\begin{equation}
    \langle P(\phi_p)\rangle= \frac{\int p_T dp_T E_p \frac{d \Pi^z (p)}{d^3p}}{\int d\phi_p p_T dp_T E_p \frac{d N(p)}{d^3p}}    .
\end{equation}
Given the above theoretical setup, we use thermal vorticity as a proxy for the spin polarization tensor ($\omega^{\mu\nu} \to \varpi^{\mu\nu}$) and use it in the thermal model discussed in more detail below.


\section{Thermal model}

\subsection{Freeze-out parametrization}

In this work, we use a thermal model that assumes simultaneous kinetic and chemical freeze-outs~\cite{Broniowski:2001we,Florkowski:2001fp}. In its standard formulation, the single freeze-out model has only four parameters: temperature ($T_{\rm f}$), baryon chemical potential ($\mu_{\rm f}$), proper time ($\tau_{\rm f}$), and system size ($r_{\rm max}$). Temperature and chemical potential are inferred from the measured ratios of hadronic abundances, while the proper time and system size are extracted from the fits to the experimental transverse-momentum spectra. As we study collisions at ultrarelativistic energies we can set $\mu=0$ ($\xi=\mu/T=0$).

As the effects of the longitudinal polarization are related to the phenomenon of elliptic flow, we adopt here an extended version of the single-freeze-out model~\cite{Broniowski:2002wp} that allows for asymmetry of the hydrodynamic flow in the transverse plane
\begin{equation}
u^\mu= \frac{1}{N} \left(\,t,\,x\sqrt{1+\delta}\,,
\,y\sqrt{1-\delta}\,,z\,\right).
\end{equation} 
This introduces an additional parameter $\delta$. For $\delta > 0$, the flow in the reaction plane(along $x$-axis) is stronger than in the out-of-plane direction(along $y$-axis), leading to the positive elliptic flow coefficients $v_2$. The parameter $N$ is determined from the normalization condition $u^\mu u_\mu = 1$, which gives
\begin{equation}
    N= \sqrt{\tau^2_{\rm f}-(x^2-y^2)\delta},
\end{equation}
where $\tau_{\rm f}$ is the proper time at freeze-out 
\begin{equation}
\tau_{\rm f}^2=t^2-(x^2+y^2+z^2).
\end{equation}

Along with the asymmetric flow profile, we introduce the asymmetry of the fireball boundary in the transverse plane. This is achieved with the boundary parametrization of the form
\begin{equation}
    x=r_{\rm max} \sqrt{1-\epsilon}\, \cos \phi,
    \qquad
    y=r_{\rm max} \sqrt{1+\epsilon}\, \sin \phi,
\end{equation}
where $\phi$ is the azimuthal angle. The parameter $\epsilon$ controls the spatial elongation of the system. For $\epsilon > 0$, the emitting source is elongated in the out-of-plane direction~\cite{Baran:2004kra}.

\begin{table}[t] 
\centering
\begin{tabular}{|c|c|c|c|c|} 
 \hline
 cent.\,(\text{\%})  & $\epsilon$ & $\delta$ & $\tau_{\rm f}$ [fm] & $r_{\textrm{max}}$\,[fm] \\ [0.5ex] 
 \hline
 0--15 & 0.055 & 0.12 & 7.666 & 6.540 \\ [0.5ex] 
 \hline
 15--30 & 0.097 & 0.26 & 6.258 & 5.417\\ [0.5ex] 
 \hline
 30--60 & 0.137 & 0.37 & 4.266 & 3.779 \\ [0.5ex] 
 \hline
\end{tabular}
\caption{Thermal model parameters used to describe the PHENIX data at $\sqrt{s_{\rm NN}}=130~{\rm GeV}$, see Ref.~\cite{Broniowski:2002wp, Baran:2004kra}.}
\label{tab.par}
\end{table}

Following earlier studies, we assume that freeze-out occurs at a fixed value of the proper time and temperature, denoted as $\tau = \tau_{\rm f}$ and $T = T_{\rm f}$. Under these conditions, a three-dimensional component of the freeze-out hypersurface, denoted as $\Delta \Sigma_\lambda$, is expressed by the formula
\begin{equation}
    \Delta \Sigma_\lambda= n_\lambda dx dy\, \tau_{\rm f}\, d\eta
\end{equation}
where
\begin{equation}
    n^\lambda = \frac{1}{\tau_{\rm f}}\left(\sqrt{\tau^2_{\rm f}+x^2+y^2}\, \cosh\,\eta,x,y,\sqrt{\tau^2_{\rm f} +x^2+y^2}\, \sinh\,\eta\right)
\end{equation}
satisfies the normalization condition $n^\lambda n_\lambda = 1$ and $\eta = \frac{1}{2} \ln[(t+z)/(t-z)]$ is the spacetime rapidity. The particle four-momentum $p^\mu$ can be parametrized in terms of the rapidity $y_p$, transverse momentum $p_T = \sqrt{p_x^2 + p_y^2}$, and the azimuthal angle $\phi_p$ as
\begin{align}
    p^\mu = \left(m_T \cosh y_p, p_T \cos{\phi_p}, p_T \sin{\phi_p},m_T \sinh y_p \right) 
\end{align}
with $m_T = \sqrt{m^2 + p_T^2}$ being the transverse mass.

In our calculations, we use previously established values of the model parameters $T_{\rm f}$, $\epsilon$, $\delta$, $\tau_{\rm f}$, and $r_{\rm max}$~\cite{Broniowski:2002wp, Baran:2004kra}, which have been fitted to the PHENIX data for three different centrality classes at the beam energy of $ \sqrt{s_{\rm NN}}= 130~$GeV, see Table~\ref{tab.par}.


\subsection{Spin polarization at freeze-out }

\begin{figure}[t]
	\includegraphics[width=\linewidth]{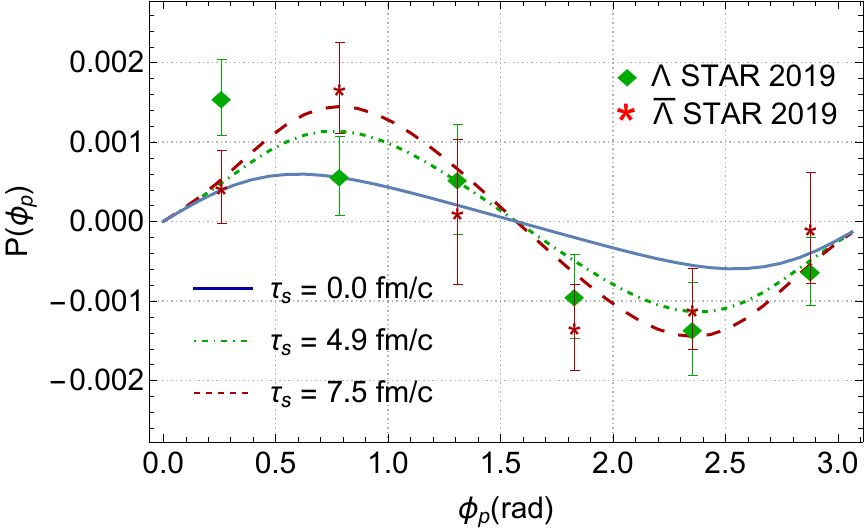}	
	\caption{(Color online) The longitudinal polarization shown as a function of the azimuthal angle $\phi_p$ for 30--60\% Au-Au collisions at $\sqrt{s_{\rm NN}} = 200$ GeV. Our theoretical results are compared with the experimental data by STAR. For more details see the description in the text.} 
	\label{fig_mom1}%
\end{figure}

The thermal vorticity can be written as a sum of two terms
\begin{equation}
    \varpi_{\mu\nu}=\underbrace{-\frac{1}{2T}(\partial_\mu u_\nu - \partial_\nu u_\mu)}_{\varpi_{\mu\nu}^I} \underbrace{+\frac{1}{2T^2}(u_\nu \partial_\mu T -u_\mu \partial_\nu T)}_{\varpi_{\mu\nu}^{II}} .
\end{equation}
Similarly, we use the following decomposition for the thermal shear
\begin{equation}
    \xi_{\mu\nu}=\underbrace{\frac{1}{2T}(\partial_\mu u_\nu + \partial_\nu u_\mu)}_{\xi_{\mu\nu}^I} \underbrace{-\frac{1}{2T^2}(u_\nu \partial_\mu T +u_\mu \partial_\nu T)}_{\xi_{\mu\nu}^{II}} .
    \label{ximunu}
\end{equation}
The derivatives of the temperature can be determined from the hydrodynamic equations
\begin{equation}
\partial^\alpha T = T\left(D\,u^\alpha- c_s^2 u^\alpha \partial \cdot u\right),
\end{equation}
where $D=u \cdot \partial$. We find that $\varpi_{\mu\nu}^{II}$ turns out to be the same as $\varpi_{\mu\nu}^{I}$. The explicit expressions for all tensor components of $\varpi_{\mu\nu}^{I}$ can be found in Ref.~\cite{Florkowski:2019voj}. The forms of $\xi_{\mu\nu}^{I}$ and $\xi_{\mu\nu}^{II}$ can be found in Ref.~\cite{Florkowski:2021xvy}. In this work, we consider $c_s^2=0.15$ \cite{HotQCD:2014kol}, which affects the polarization results through Eq.~\eqref{ximunu}. The derivatives of thermal vorticity ($\partial_\lambda \varpi_{\mu\nu}$) are listed in \ref{appB}.


\section{Results and Discussion}

In this section, we present our numerical results for the $p_T$-integrated longitudinal component of the mean Pauli-Lubanski four-vector, which characterizes the longitudinal spin polarization of the $\Lambda$ and $\overline{\Lambda}$ hyperons. The model parameters utilized in the calculations are provided in Table~\ref{tab.par}. These parameters were previously fitted to describe the PHENIX data at the beam energy of $\sqrt{s_{\rm NN}} = 130$ GeV for three centrality classes: $0$--$15\%$, $15$--$30\%$, and $30$--$60\%$, at freeze-out temperature $T_{\rm f} = 0.165$ GeV.

As our primary interest lies in the longitudinal polarization, there is no need to boost the quantity $\Pi_z(p_x, p_y)$ back to the particle's rest frame due to its invariance under transverse boosts. Additionally, to align with experimental practices, we solely focus on the central rapidity region ($y_p = 0$). We perform the transverse momentum ($p_T$) integral in the range from 0 to 3 GeV, and we have verified that extending the range to the experimental momentum range of 0.5--6.0 GeV does not significantly alter the results.

\begin{figure}[t]
	\includegraphics[width=\linewidth]{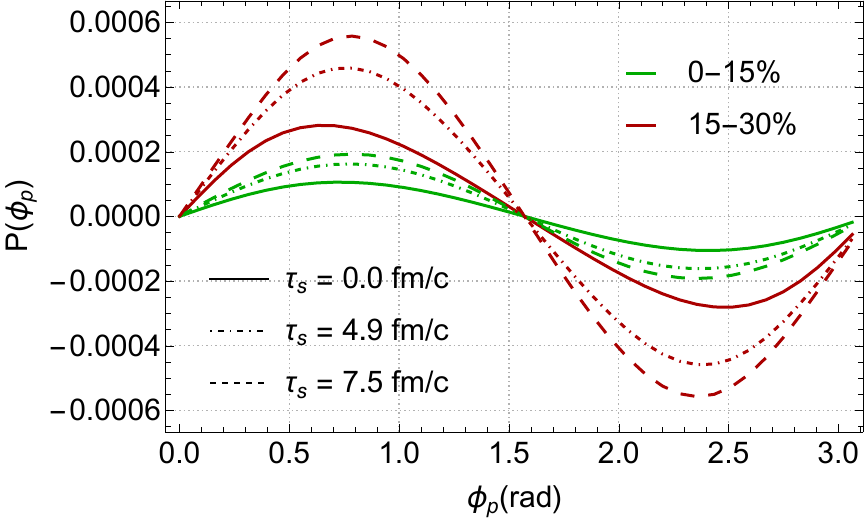}
	\caption{(Color online) Our predictions for the Au-Au collisions at $\sqrt{s_{\rm NN}} = 200$~GeV for centralities $0-15\%$ and $15-30\%$. Dashed-dotted lines for the relaxation time 4.9 fm/$c$ and dashed lines for the relaxation time 7.5 fm/$c$. For comparison, we also show the case without dissipative corrections, represented by solid lines. The Green lines represent 0-15\% centrality and, the red lines represent 0-30\% centrality.} 
	\label{fig_mom2}%
\end{figure}

The magnitude of dissipative corrections depends on the value of the spin relaxation time $\tau_s$. We observe that the inclusion of dissipative correction leads to an increase in the magnitude of longitudinal polarization. This feature may be attributed to the fact that dissipation results in increased conversion of fluid angular momentum to spin polarization. To determine the value of $\tau_s$, we have performed a reduced chi-squared test. We find that the optimal description of the data for $\overline{\Lambda}$ hyperons is achieved with $\tau_s=7.5$ fm/$c$, corresponding to the reduced chi-squared value $\chi^2_{\rm r}=0.6$. On the other hand, for $\Lambda$ hyperons, we find good agreement with $\tau_s=4.9$ fm/$c$ corresponding to $\chi^2_{\rm r}=1.5$. These values of the spin relaxation times are in agreement with those obtained in Refs.~\cite{Kapusta:2019sad, Ayala:2020ndx, Hidaka:2023oze, Wagner:2024fhf}.

In Fig.~\ref{fig_mom1} we show the longitudinal polarization as a function of the azimuthal angle $\phi_p$  for 30--60\% Au-Au collisions at $\sqrt{s_{\rm NN}} = 200$ GeV. Experimental data points are taken from \cite{STAR:2019erd}, and conversion from $\langle\cos\theta^*_p\rangle$ to polarization helicity (PH) is conducted using $\alpha_H = 0.732$ \cite{ParticleDataGroup:2020ssz}. Error bars represent the combined statistical and systematic uncertainties. The solid line represents our theoretical result without dissipative corrections, the dashed line corresponds to the dissipative case with $\tau_s= 7.5 ~ {\rm fm}/c$, while the dashed-dotted line corresponds to $\tau_s=4.9~ {\rm fm}/c$. In Fig.~\ref{fig_mom2} we present our theoretical predictions for two other centrality classes for the same colliding system. 

\section{Summary and conclusions}

In this study, we addressed the issue of longitudinal spin polarization of $\Lambda$ hyperons generated in relativistic heavy-ion collisions. We employed a relativistic kinetic-theory framework, incorporating classical treatment of spin degrees of freedom, combined with the freeze-out parametrization utilized in prior investigations. By employing kinetic theory, we were able to include dissipative corrections, such as those arising from thermal shear and gradients of thermal vorticity, into the Pauli-Lubanski vector, which determines spin polarization. This allowed for direct comparison with experimental data. Similar to previous studies, we found that a comprehensive explanation of the data necessitated additional assumptions. Specifically, we employed projected thermal vorticity as a proxy to spin polarization tensor. The spin relaxation time was fitted to match the experimental data through a reduced chi-squared minimization procedure. The estimated spin relaxation time was found to be approximately $5$ fm/$c$, which is consistent with earlier estimates~\cite{Kapusta:2019sad, Ayala:2020ndx, Hidaka:2023oze, Wagner:2024fhf}. From this analysis, we conclude that dissipative corrections to spin tensor play an important role in explaining the observed longitudinal polarization in heavy-ion collisions.

\section*{Acknowledgements}

The authors would like to thank Sourav Dey for useful discussions. S. Banerjee acknowledges the kind hospitality of Jagiellonian University, where part of this work was carried out. This work was supported in part by the Polish National Science Centre Grants No. 2022/47/B/ST2/01372 (W.F. and S. Banerjee) and No. 2018/30/E/ST2/00432 (R.R.).

\appendix

\section{Pauli-Lubanski vector with dissipative corrections}
\label{appA}

In this section, we outline the steps to evaluate Eq.~\eqref{PL-KT} starting from the kinetic theory expression of the non-equilibrium correction to phase-space distribution function, given by,
%
%
\begin{align}
    \delta f_s^{\pm} &= -\frac{\tau_s}{u \cdot p} e^{\pm \xi - \beta \cdot p} \left[ - p^\lambda p^\mu \left(\partial_\mu\, \beta_\lambda\right) \left(1+\frac{1}{2}s^{\alpha \beta} \omega_{\alpha \beta}\right) \right. \nonumber\\ 
    &\qquad+ \left.\frac{1}{2} p^\mu s^{\alpha \beta} \left(\partial_\mu \omega_{\alpha \beta} \right) \right]
\end{align}
Using this we can determine the off-equilibrium correction to spin tensor as,
\begin{align}
    \delta S^{\lambda,\mu \nu} &=\int dP dS p^\lambda s^{\mu \nu} \left(\delta f_s^{+} + \delta f_s^{-} \right) \nonumber\\
    &= -2 \cosh \xi \int dP dS p^\lambda s^{\mu \nu} \frac{\tau_s}{u \cdot p} e^{-\beta \cdot p} \nonumber\\
    &\times \left[- p^\rho p^\sigma \left(\partial_\sigma\, \beta_\rho\right) \left(1+\frac{1}{2}s^{\alpha \beta} \omega_{\alpha \beta}\right) \right. \left. + \frac{1}{2} p^\rho s^{\alpha \beta}(\partial_\rho \omega_{\alpha \beta}) \right] \nonumber\\
    &= \tau_s \cosh \xi\, \omega_{\alpha \beta} \left( \partial_\sigma\, \beta_\rho\right) \!\int \frac{dP dS}{\left(u \cdot p\right)} p^\lambda p^\rho p^\sigma s^{\mu \nu} s^{\alpha \beta}\, e^{- \beta \cdot p} \nonumber\\
    &- \tau_s \cosh\xi\, (\partial_\rho \omega_{\alpha \beta}) \!\int \frac{dP dS}{\left(u\cdot p\right)} p^\lambda p^\rho s^{\mu \nu} s^{\alpha \beta}\, e^{- \beta \cdot p}. \label{app:dS-1}
\end{align}
where we have used the result $\int dS s^{\mu \nu} = 0$ as derived in Refs.~\cite{Florkowski:2018fap, Bhadury:2020cop} and the integral measures are defined as,
\begin{align}
    dP &\equiv  \frac{d^3 p}{\left(2\pi\right)^3 p^0}, \label{app:dP}\\
    dS &\equiv  \frac{4 m}{3 \pi}\, d^4s~ \delta \left(s \cdot s + \frac{3}{4}\right) \delta \left( p\cdot s\right).\label{app:dS}
\end{align}
Furthermore, using the relations \cite{DeGroot:1980dk, Florkowski:2018fap, Bhadury:2020cop},
\begin{align}
    \int dS s^{\mu \nu} s^{\alpha \beta} &= -\frac{1}{2m^2} \epsilon^{\mu \nu \rho \sigma} \epsilon^{\alpha \beta \gamma \delta} p_\rho p_\gamma g_{\sigma \delta}, \label{app:i-ss}\\
    \omega_{\alpha \beta} \epsilon^{\mu \nu \rho \sigma} \epsilon^{\alpha \beta \gamma \delta} p_\rho p_\gamma g_{\sigma \delta} &= -\left(2m^2 \omega^{\mu \nu} + 4 p_\rho p^{[\mu} \omega^{\nu] \rho} \right), \label{app:identity1}
\end{align}
we can carry out the integration over spin to obtain,
\begin{align}
    \delta S^{\lambda,\mu \nu} &= \frac{\tau_s \cosh\xi}{m^2} \int \frac{dP}{\left(u\cdot p\right)} p^\lambda e^{- \beta \cdot p} \nonumber\\
    &\times \left[ p^\rho p^\sigma \left(\partial_\sigma\, \beta_\rho \right) \left(m^2 \omega^{\mu\nu} + 2 p_\alpha p^{[\mu} \omega^{\nu]\, \alpha}\right)\right. \nonumber\\
    &\left.- m^2 p^{\,\rho} \left(\partial_\rho \omega^{\mu\nu}\right) - 2 p^{\,\rho} p_\alpha p^{[\mu } \left(\partial_\rho \omega^{\nu]\, \alpha} \right) \right],
\end{align}
where the square brackets imply anti-symmetric combination i.e., $A^{[\mu} B^{\nu]} = \left(A^\mu B^\nu - A^\nu B^\mu\right)/2$. The Pauli-Lubanski vector in Eq.~\eqref{eq:PL1} can then be expressed as
\begin{align}
    &E_p \frac{d \Pi_\alpha (x,p)}{d^3p} = - \frac{1}{2} \epsilon_{\alpha \mu \nu \beta} \Delta \Sigma_\lambda E_p\frac{dS^{\lambda,\mu \nu}}{d^3p}\frac{p^\beta}{m} \nonumber\\
    &= - \frac{\cosh\xi}{2 m\, (2\pi)^3} e^{-\beta \cdot p} \epsilon_{\alpha \mu \nu \beta } \,\Delta \Sigma \cdot p \nonumber\\
    &\qquad\times \bigg[ p^\beta \omega^{\mu \nu} + \frac{\tau_s\, p^\beta}{\left(u \cdot p\right)} \left\{ p^\rho p^\sigma \omega^{\mu \nu} \left(\partial_\sigma \beta_\rho\right) - p^\rho \left(\partial_\rho \omega^{\mu \nu}\right) \right\} \bigg] \nonumber\\
    &= - \frac{\cosh\xi}{2 m\, (2\pi)^3} e^{-\beta \cdot p} \epsilon_{\alpha \mu \nu \beta }  p^\beta\,\Delta \Sigma \cdot p  \, \left[ \left(1 + \chi \right) \omega^{\mu \nu } - \frac{\tau_s}{\left(u \cdot p\right)} \; p^\rho \left(\partial_\rho \omega^{\mu \nu} \right) \right]
\end{align}
Integrating over the spacetime volume we can write,
\begin{align}
    E_p \frac{d \Pi_\alpha (p)}{d^3p} &= - \frac{\cosh(\xi)}{2 m\, (2\pi)^3} \!\int\! \Delta \Sigma \cdot p\, \epsilon_{\alpha \mu \nu \beta} p^\beta\, e^{-\beta \cdot p} \nonumber\\
    &\quad \times \left[ \left(1 + \chi \right) \omega^{\mu \nu}- \frac{\tau_s}{\left(u \cdot p\right)}\;  p^\rho \left(\partial_\rho \omega^{\mu \nu}\right) \right]
\end{align}
with,
\begin{align}
    \chi &= \frac{\tau_s}{\left(u \cdot p\right)}  p^{\,\rho} p^\sigma \left(\partial_\sigma \beta_\rho\right)
    \,=\, \frac{\tau_s\, p^{\,\rho} p^\sigma \xi_{\sigma \rho}}{ \left(u \cdot p\right)}. \label{chi-def}
\end{align}

\section{Derivative of Thermal Vorticity Components}
\label{appB}
Here we list the derivatives of the ``magnetic like" components of the thermal vorticity.
\begin{align}
   \partial_0 \varpi^{12} &=\! \frac{-3txy}{T_{\rm f} N^5} \left[\! \sqrt{1 \!-\! \delta } -\! \sqrt{1 \!+\! \delta} + \delta \left(\sqrt{1 \!-\! \delta} +\! \sqrt{1 \!+\! \delta} \right) \right] \label{d0O12},\\
   \partial_1 \varpi^{12} &= \frac{y}{T_{\rm f} N^5} \Big[\sqrt{1 \!- \delta } -\! \sqrt{1 \!+\delta} + \delta \left(\sqrt{1 \!- \delta} +\! \sqrt{1 \!+ \delta} \right) \Big] \nonumber\\
   &\quad \times \left[t^2+2 (1+\delta ) x^2-(1-\delta)y^2-z^2\right] \label{d1O12}, \\
   \partial_2\varpi^{12} &=\! \frac{x}{T_{\rm f} N^5} \left[\!\sqrt{1 \!-\! \delta } -\! \sqrt{1 \!+\! \delta} + \delta \left(\sqrt{1 \!-\! \delta} +\! \sqrt{1 \!+\! \delta} \right) \right] \nonumber\\
   &\quad \times \left[ t^2-(1+\delta) x^2+2 (1-\delta)  y^2-z^2\right] \label{d2O12}, \\
   \partial_3\varpi^{12} &=\! \frac{3xyz}{T_{\rm f} N^5} \left[\! \sqrt{1 -\! \delta } -\! \sqrt{1 \!+\! \delta} +\! \delta \left(\sqrt{1 \!-\! \delta} + \sqrt{1 \!+\! \delta}\right)\right] \label{d3O12}, \\
   \partial_0\varpi^{13} &= - \frac{3txz}{T_{\rm f} N^5}\left[1+\delta -\sqrt{1+\delta}\right] \label{d0O13}, \\
   \partial_1\varpi^{13} &= \frac{z}{T_{\rm f} N^5} \left[1+\delta -\sqrt{1+\delta}\right] \nonumber\\
   &\quad \times\left[ t^2+2 (1+\delta) x^2-(1-\delta)y^2-z^2\right] \label{d1O13}, \\
   \partial_2\varpi^{13} &= \frac{3xyz}{T_{\rm f} N^5} \left[1+\delta -\sqrt{1+\delta}\right] (1-\delta)\label{d2O13}, \\
   \partial_3\varpi^{13} &= \frac{x}{T_{\rm f} N^5} \left[1+\delta -\sqrt{1+\delta}\right] \nonumber\\
   &\quad\times \left[t^2-(1+\delta) x^2-(1-\delta)y^2+2 z^2\right] \label{d3O13}, \\
   \partial_0\varpi^{23} &= \frac{-3tyz}{T_{\rm f} N^5}\left[1-\delta-\sqrt{1-\delta }\right] \label{d0O23}, \\
   \partial_1\varpi^{23} &= \frac{3xyz}{T_{\rm f} N^5} \left[1-\delta-\sqrt{1-\delta }\right] (1+\delta)\label{d1O23}, \\
   \partial_2\varpi^{23} &= \frac{z}{T_{\rm f} N^5}\left[1-\delta -\sqrt{1-\delta }\right] \nonumber\\
   &\quad \times\left[t^2-(1+\delta) x^2+2 (1-\delta)y^2-z^2\right] \label{d2O23}, \\
   \partial_3\varpi^{23} &=\frac{y}{T_{\rm f} N^5}\left[1-\delta -\sqrt{1-\delta }\right] \nonumber\\
   &\quad \times\left[t^2 - (1 + \delta) x^2 - (1 - \delta) y^2 + 2 z^2\right] \label{d3O23}.
\end{align}


\bibliographystyle{elsarticle-num} 
\bibliography{ref}






\end{document}